\documentclass[11pt]{article}

\usepackage{color, amsmath, amssymb, amsbsy, amsthm, graphicx, bbm, amsfonts, bm, dsfont, courier, subfig}
\usepackage[toc,page]{appendix}
\usepackage{setspace}
\usepackage{booktabs}
\usepackage[flushleft]{threeparttable}
\usepackage{latexsym}
\usepackage[dvipsnames]{xcolor}
\usepackage{comment}
\usepackage{wrapfig}
\usepackage{graphicx}
\usepackage{graphics}
\usepackage{algorithm}
\usepackage{algorithmic}
\usepackage{makecell}
\usepackage[colorlinks,allcolors=blue]{hyperref}
\usepackage[numbers]{natbib}  
\usepackage{tikz}
\allowdisplaybreaks
\usepackage[figuresright]{rotating}
\usepackage{soul}
\usepackage{threeparttable,booktabs}
\usepackage{etoolbox}
\appto\TPTnoteSettings{\footnotesize}
\usepackage{svg} 
\usepackage{tablefootnote}
\usepackage{authblk}
\usepackage{enumitem}

\newtheoremstyle{theoremstyle}
{\topsep} 
{\topsep} 
{\itshape} 
{} 
{} 
{} 
{.5em} 
{\color{black}\ifthenelse{\equal{#3}{}}{{\bfseries #1 #2}}{{\bfseries #1 #2 (#3)}}}
\newtheoremstyle{theoremstylealt}
{\topsep} 
{\topsep} 
{\itshape} 
{} 
{} 
{} 
{.5em} 
{\color{black}\ifthenelse{\equal{#3}{}}{{\bfseries #1 #2$^\prime$}}{{\bfseries #1 #2$^\prime$ (#3)}}}
\newtheoremstyle{examplestyle}
{\topsep} 
{\topsep} 
{} 
{} 
{} 
{} 
{.5em} 
{\color{black}\ifthenelse{\equal{#3}{}}{{\bfseries #1 #2}}{{\bfseries #1 #2 (#3)}}}
\theoremstyle{theoremstyle}\newtheorem{thm}{Theorem}
\theoremstyle{theoremstylealt}
\theoremstyle{theoremstyle}     
\theoremstyle{theoremstyle}  
\theoremstyle{theoremstyle}        
\theoremstyle{theoremstyle}
\theoremstyle{theoremstyle}\newtheorem{assumption}{Assumption}
\theoremstyle{theoremstylealt}
\theoremstyle{theoremstyle}
\theoremstyle{theoremstyle}

\theoremstyle{theoremstyle}

\theoremstyle{examplestyle}
\theoremstyle{examplestyle}
\theoremstyle{examplestyle}

\renewcommand{\epsilon}{\varepsilon}

\def \hat{\widehat}

\usepackage{setspace}
\usepackage{array}
\newcolumntype{H}{>{\setbox0=\hbox\bgroup}c<{\egroup}@{}}
\definecolor{applegreen}{rgb}{0.55, 0.71, 0.0}
\begin{document}

	\onehalfspacing
	\title{Rerandomized Egger estimator in two-sample summary-data Mendelian randomization}

\author[1]{Youpeng Su}
\author[2]{Yilei Ma}
\author[1]{Ping Yin}
\author[1, ]{Peng Wang \thanks{Correspondence: \href{mailto:pengwang\_stat@hust.edu.cn}{pengwang\_stat@hust.edu.cn}}}

\affil[1]{Department of Epidemiology and Biostatistics, School of Public Health, Tongji Medical College, Huazhong University of Science and Technology}
\affil[2]{Tongji Hospital, Tongji Medical College, Huazhong University of Science and Technology}

\maketitle

\begin{abstract}
In two-sample Mendelian randomization (MR), 
MR-Egger regression remains one of the most widely 
used methods because of its robustness to directional pleiotropy. 
However, it ignores the measurement error in the genetic associations with the exposure 
and is therefore susceptible to measurement-error bias. 
A stringent threshold for selecting instrumental variables (IVs) 
can reduce this bias but often retains too few instruments, inflating variance; 
by contrast, 
more liberal thresholds retain more instruments 
but can exacerbate winner's curse bias
when the same exposure GWAS is used 
for both IV selection and effect estimation.
To address these challenges, 
we extend the recently developed rerandomized inverse variance weighted 
(RIVW) estimator to accommodate directional pleiotropy 
and propose the rerandomized Egger (REgger) estimator.
REgger simultaneously corrects measurement-error bias and winner's curse bias 
while allowing the inclusion of more moderate-effect IVs to improve precision.
Under appropriate conditions, 
we show that the REgger estimator is asymptotically normal 
and admits a closed-form variance estimator.
Simulation studies and real-data applications 
demonstrate that, in the presence of directional pleiotropy, 
REgger generally outperforms RIVW in both bias and precision.
  \par\vspace{1ex} 
  \noindent \textbf{Keywords:} Rerandomized Egger estimator; 
                               directional pleiotropy; 
                               winner's curse;
                               measurement error.	
\end{abstract}

\clearpage
\section{Introduction}\label{Section-introduction}
\begin{figure}[htbp]
  \centering
  \includegraphics[width=0.4\textwidth]{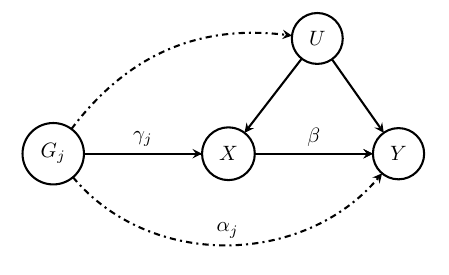}
  \caption{Directed acyclic graph illustrating the core instrumental variable assumptions 
  for the genetic variant $G_j$ (solid arrows), together with potential violations (dashed arrows). 
  $U$ denotes the unmeasured confounder,
   and $\gamma_j$, $\beta$, and $\alpha_j$ denote the $G_j\rightarrow X$, $X\rightarrow Y$, 
   and direct $G_j \rightarrow Y$ effects, respectively.}
  \label{Fig_intor_1}
\end{figure}

Establishing causal relationships between exposures and health outcomes is a central goal in epidemiology. 
Mendelian randomization (MR) uses genetic variants as natural instrumental variables (IVs) 
to strengthen causal inference from observational data
by exploiting the random allocation of alleles at conception, 
thereby mimicking certain aspects of randomization.
In two-sample summary-data MR, 
summary statistics for single nucleotide polymorphism (SNP)--exposure 
and SNP--outcome associations are obtained 
from separate genome-wide association studies (GWASs), 
enabling scalable causal investigation across a wide range of phenotypes 
without requiring individual-level data \cite{burgess2013mendelian,pierce2013efficient,sanderson2022mendelian}. 
A valid genetic instrument relies on three core IV assumptions,
illustrated in Figure \ref{Fig_intor_1} \cite{didelez2007mendelian,sanderson2022mendelian}:

\begin{enumerate}[ 
   label=A-\Roman*., 
   itemsep=-3pt,      
    align=left,        
    leftmargin=*,      
    labelwidth=1em,    
    labelsep=0em     
]
    \item The IV must be associated with the exposure $X$ (relevance assumption);
    \item The IV must be independent of any confounders $U$ (independence assumption);
    \item The IV must influence the outcome $Y$ only through the exposure $X$ (exclusion-restriction assumption).
\end{enumerate}

A key threat to the validity of MR is horizontal pleiotropy,
whereby genetic variants influence the outcome through pathways other than the exposure, 
violating the exclusion-restriction assumption \cite{hemani2018evaluating,solovieff2013pleiotropy}.
When pleiotropic effects display a systematic tendency in one direction, 
directional pleiotropy arises, 
under which the commonly used inverse-variance weighted (IVW) estimator can be biased \cite{burgess2013mendelian}.
To address directional pleiotropy, 
MR-Egger regression was developed by introducing an intercept term, 
and under the InSIDE (Instrument Strength Independent of Direct Effect) assumption,
the MR-Egger estimator is theoretically consistent \cite{bowden2015mendelian,burgess2017interpreting}.
Consequently, MR-Egger has become one of the most widely used methods in MR studies 
and is frequently employed as a sensitivity analysis alongside IVW.
Despite its theoretical appeal and widespread use, 
MR-Egger suffers from practical limitations in modern GWAS settings 
that create a difficult trade-off 
between bias control and estimation precision.
In particular, 
MR-Egger ignores measurement error in SNP--exposure associations,
which can bias the causal estimate toward the null \cite{bowden2016assessing,burgess2017interpreting,bowden2019improving,lin2022practical}.
To mitigate this measurement-error bias, 
a natural approach is to apply a stringent threshold for IV selection, 
thereby restricting the analysis to stronger variants.
However, stringent thresholding often retains only a limited number of IVs, 
which can substantially inflate the variance of the causal estimator.
Conversely, more liberal thresholds allow the inclusion of additional IVs 
and may improve precision, 
but when the same exposure GWAS is used for both IV selection and effect estimation, 
they can exacerbate winner's curse bias \cite{ma2023breaking,jiang2023empirical}.
Thus, in practice, MR-Egger faces a nontrivial trade-off 
between bias control and estimation precision.

Several methodological advances have been proposed to address these issues.
In the IVW framework, the debiased IVW (dIVW) estimator corrects 
measurement-error bias by adjusting the ratio denominator \cite{ye2021debiased},
and the rerandomized IVW (RIVW) estimator further mitigates winner's curse 
through random IV selection \cite{ma2023breaking},
thereby improving precision while controlling bias.
However, both dIVW and RIVW are limited to balanced pleiotropy 
and do not directly extend to accommodate directional pleiotropy.
Existing refinements of MR-Egger are also limited, 
as they do not jointly address measurement-error bias 
and winner's curse bias \cite{bowden2019improving,lin2022practical}.
Importantly, the MR-Egger slope estimator can be expressed as a ratio estimator, 
which shares a structural similarity with the IVW estimator.
This key insight makes it possible to adapt 
the construction strategy of RIVW to the Egger framework, 
thereby enabling a unified solution that corrects both measurement-error bias and winner's curse bias 
in the context of directional pleiotropy.

While MR-Egger inference may be affected by 
additional practical considerations, such as SNP coding, 
the present study is designed to comprehensively 
address the primary sources of bias in causal effect estimation. 
Specifically, we use a fixed minor-allele-based coding scheme 
to avoid data-driven all-positive recoding, 
and examine SNP coding sensitivity in our simulation studies.
The Egger-intercept diagnostic problem is a distinct and complementary topic,
addressed in dedicated companion work by Ma et al.~\cite{ma2026modified}.
The present study does not develop, evaluate, or apply an intercept-based pleiotropy test.
It instead focuses on reliable estimation of the causal effect under directional pleiotropy
through development and validation of the REgger estimator.

In this paper, 
we propose the rerandomized Egger (REgger) estimator, 
which extends the RIVW framework to accommodate directional pleiotropy 
and simultaneously addresses measurement-error bias and winner's curse bias.
REgger combines the Egger regression framework 
with rerandomized IV selection and Rao--Blackwellized 
SNP--exposure effect estimates,
enabling the inclusion of more moderate-effect IVs to improve precision 
while maintaining robustness to directional pleiotropy.
Under appropriate conditions, 
we establish the asymptotic normality of REgger 
and derive a closed-form variance estimator. 
Through extensive simulation studies and real-data applications, 
we demonstrate that, 
in the presence of directional pleiotropy, 
REgger generally outperforms RIVW in terms of both bias and precision.

\section{Assumptions and Notation}
During the data-cleaning stage, 
millions of SNPs are typically decorrelated using linkage disequilibrium (LD) clumping or pruning (e.g., via PLINK)\cite{hemani2018mr,purcell2007plink}. 
After this step, we assume that there are $p$ (approximately) independent genetic variants $G_1,\cdots, G_p$, 
where each $G_j\in\{0,1,2\}$ denotes the allele count for SNP $j$ in an individual.
Following the two-sample summary-data MR literature \cite{ye2021debiased,ma2023breaking,zhao2020statistical}, 
we adopt the linear structural models:
\begin{align}
  X=&\sum_{j=1}^{p}{\gamma_jG_j}+U+E_X, \label{Eqa_exp} \\
  Y=&\sum_{j=1}^{p}{\alpha_jG_j}+\beta X+U+E_Y. \label{Equ_out}
\end{align} 
Here, $\gamma_j$ denotes the genetic effect of $G_j$ on the exposure, 
$\alpha_j$ denotes the direct (pleiotropic) effect of $G_j$ on the outcome that is not mediated through $X$,
and $\beta$ is the causal parameter of interest. 
The term $U$ denotes an unmeasured (aggregated) confounder, and $E_X$ and $E_Y$ are independent random errors, 
assumed to be independent of $\left(G_1,\cdots,G_p,U\right)$. 
The total genetic effect of $G_j$ on the outcome is therefore given by
$\Gamma_j=\alpha_j+\beta\gamma_j$.
Throughout, we work under the InSIDE assumption,
meaning that the pleiotropic effects $\alpha_j$ are independent of
the instrument strengths $\gamma_j$ across SNPs.

In practice, individual-level data are often unavailable due to privacy or logistical constraints. 
Consequently, two-sample summary-data MR is widely used, 
relying on SNP--exposure and SNP--outcome association estimates obtained from independent GWASs \cite{sanderson2022mendelian,pierce2013efficient}. 
For each SNP $j$, the exposure GWAS provides the marginal regression coefficient 
${\hat{\gamma}}_j$, along with standard error $\sigma_{Xj}$, 
and the outcome GWAS provides the corresponding quantities $({\hat{\Gamma}}_j,\sigma_{Yj})$. 
To facilitate asymptotic analysis, we introduce the following assumptions.
\begin{assumption}
  \label{Assume1} 
  The sample sizes of the exposure and outcome GWASs, 
  denoted by $n_X$ and $n_Y$, respectively,
   satisfy $\min(n_X,n_Y)\to\infty$, and $n_X$ and $n_Y$ are of the same order. 
  Meanwhile, the number of (approximately) independent SNPs satisfies $p\rightarrow\infty$.
\end{assumption}

\begin{assumption} 
  \label{Assume2}
  The set of $2p$ estimated genetic associations $\left\{{\hat{\gamma}}_j,{\hat{\Gamma}}_j\right\}_{j=1}^p$ are mutually independent. 
  For each SNP $j$,
  $$\begin{bmatrix}{\hat{\Gamma}}_j\\{\hat{\gamma}}_j\end{bmatrix}\sim \mathcal{N} \left(\begin{bmatrix}\Gamma_j\\\gamma_j\end{bmatrix},\begin{bmatrix}\sigma_{Yj}^2&0\\0&\sigma_{Xj}^2\end{bmatrix}\right)$$
  with known variances $\sigma_{Yj}^2$ and $\sigma_{Xj}^2$. 
  Furthermore, for $n=\min(n_X,n_Y)$, the quantities $n\sigma_{Yj}^2$ 
  and $n\sigma_{Xj}^2$ are bounded and bounded away from zero. 
\end{assumption}

Assumption \ref{Assume1} is well supported by modern GWASs, which typically involve tens to hundreds of thousands of 
participants, with a few thousand SNPs remaining for analysis after the decorrelation step \cite{ye2021debiased,uffelmann2021genome}.
The two-sample MR design ensures independence between the SNP--exposure estimates 
$\left({\hat{\gamma}}_1,\cdots,{\hat{\gamma}}_p\right)$ and 
the SNP--outcome estimates $({\hat{\Gamma}}_1,\cdots,{\hat{\Gamma}}_p)$ by construction. 
Approximate independence within each set is justified by two considerations:
first, LD clumping or pruning greatly reduces genetic correlations among SNPs; 
second, each SNP explains only a tiny fraction of the variance in $X$ or $Y$, 
so residual dependencies among their estimates are negligible.
The approximate normality and known-variance assumptions are standard 
in the MR literature 
given the large GWAS sample sizes \cite{ye2021debiased,ma2023breaking}. 
For simplicity, we further assume that $\sigma_{Xj}^2$ and $\sigma_{Yj}^2$ are of the 
same order as $1/n$. 

To facilitate discussion, we define the average IV strength for $p$ instruments as 
$$\kappa=p^{-1}\sum_{j=1}^{p}{\sigma_{Xj}^{-2}\gamma_j^2},$$
which summarizes the overall strength of the SNP--exposure associations. 
Throughout, we use $\xrightarrow{P}$ and $\xrightarrow{D}$ to denote convergence in probability and distribution, 
respectively. 
For two random variables, $A$ and $B$, we write $A=O_p\left(B\right)$ if $A/B$ is asymptotically bounded 
in probability, and $A=o_p\left(B\right)$ if $A/B\xrightarrow{P}0$.

\section{Method}\label{Method}

\subsection{Measurement-error Bias in MR-Egger}
MR-Egger regression extends the IVW framework by incorporating an intercept term
 to account for average directional pleiotropy \cite{bowden2015mendelian,burgess2017interpreting}. 
Specifically,
$${\hat{\Gamma}}_j=\mu_\alpha+\beta{\hat{\gamma}}_j+\varepsilon_{Ej}, \ \ \varepsilon_{Ej}\sim N\left(0,\sigma_E^2\sigma_{Yj}^2\right),$$
where $\varepsilon_{Ej}$ is a random error term
and $\sigma_E^2\ge 1$ is an unknown overdispersion parameter. 
The intercept $\mu_\alpha$ represents the mean pleiotropic effect across instruments, while the slope
$\beta$ identifies the causal effect under the InSIDE assumption.
To study the large-sample behavior of MR-Egger and REgger, we impose the following assumption.
\begin{assumption}\label{InSIDE_Assumption}
	For all $j$, $\Gamma_j=\alpha_j+\beta\gamma_j$. 
	Moreover, $\alpha_j$ is independent of $\gamma_j$, 
	with ${E}\left(\alpha_j\right)=\mu_\alpha$ and ${\rm Var}\left(\alpha_j\right)=\tau_\alpha^2$,
	and $\alpha_j^2=O\left(\kappa/n\right)$ uniformly over $j$.
\end{assumption}
Assumption \ref{InSIDE_Assumption} 
formalizes the InSIDE condition 
and ensures identifiability of the causal slope $\beta$ 
in Egger-type regression. 
Using $\sigma_{Yj}^2 = \Theta(1/n)$ from Assumption \ref{Assume2},
the bound  $\alpha_j^2=O\left(\kappa/n\right)$ 
can equivalently be written as
$\alpha_j^2\sigma_{Yj}^{-2}=O\left(\kappa\right)$, 
meaning that the magnitude of the pleiotropic effect 
on the outcome is of no larger order 
than the average instrument strength for the exposure.

The slope is estimated by weighted least squares with weights $\sigma_{Yj}^{-2}$, yielding
$${\hat{\beta}}_{\rm Egger}=\frac{{\hat{\theta}}_1}{{\hat{\theta}}_2},$$
where
$${\hat{\theta}}_1
 =\Big(\sum_{j=1}^{p}\sigma_{Yj}^{-2}\Big)\Big(\sum_{j=1}^{p}{\sigma_{Yj}^{-2}{\hat{\gamma}}_j{\hat{\Gamma}}_j}\Big)
  -\Big(\sum_{j=1}^{p}{\sigma_{Yj}^{-2}{\hat{\gamma}}_j}\Big)\Big(\sum_{j=1}^{p}{\sigma_{Yj}^{-2}{\hat{\Gamma}}_j}\Big)$$
represents the weighted sample covariance 
between the SNP--exposure and SNP--outcome associations
and
$$\hat{\theta}_2=\Big(\sum_{j=1}^{p}\sigma_{Yj}^{-2}\Big)\Big(\sum_{j=1}^{p}{\sigma_{Yj}^{-2}{\hat{\gamma}}_j^2}\Big)-\Big(\sum_{j=1}^{p}{\sigma_{Yj}^{-2}{\hat{\gamma}}_j}\Big)^2$$
represents the weighted sample variance 
of the SNP--exposure associations. 

The attenuation of MR-Egger caused by measurement error in
 $\hat{\gamma}_j$ 
has been discussed previously by Bowden et al. (2016) 
from a regression-based perspective. 
Here we revisit the same issue through the ratio representation above, 
which more directly reveals the source of denominator inflation and, 
in turn, motivates the construction of the proposed REgger estimator.

To study the impact of measurement error, define the corresponding population quantities
$$
  \theta_2=\Big(\sum_{j=1}^{p}\sigma_{Yj}^{-2}\Big)\Big(\sum_{j=1}^{p}{\sigma_{Yj}^{-2}\gamma_j^2}\Big)-\Big(\sum_{j=1}^{p}{\sigma_{Yj}^{-2}\gamma_j}\Big)^2>0,
$$
and
$$
  \Delta=\Big(\sum_{j=1}^{p}\sigma_{Yj}^{-2}\Big)\Big(\sum_{j=1}^{p}{\sigma_{Yj}^{-2}\sigma_{Xj}^2}\Big)-\sum_{j=1}^{p}{\sigma_{Yj}^{-4}\sigma_{Xj}^2}>0.
$$
Under Assumption \ref{Assume2}, 
and using the InSIDE condition in Assumption \ref{InSIDE_Assumption} for 
$E(\hat{\theta}_1)$, a direct calculation shows that
$$
E({\hat{\theta}}_1)=\beta\theta_2, \quad E({\hat{\theta}}_2)=\theta_2+\Delta,
$$
and hence
$$E\big({\hat{\beta}}_{\rm Egger}\big)\approx\frac{E({\hat{\theta}}_1)}{E({\hat{\theta}}_2)}
 =\frac{\beta\theta_2}{\theta_2+\Delta}
=\frac{\beta}{1+\Delta/\theta_2}.
$$
Thus, the relative bias of ${\hat{\beta}}_{\rm Egger}$ is governed by the ratio $\Delta/\theta_2$. 
When measurement error in $\hat{\gamma}_j$ is substantial, 
$\Delta$ becomes large and ${\hat{\beta}}_{\rm Egger}$ is attenuated toward the null.
Consistency would require $$\Delta/\theta_2\rightarrow0,$$
which generally entails infinitely strong instruments ($\kappa\rightarrow\infty$),
a condition rarely met in practice.

Because ${\hat{\theta}}_2$ 
depends on the empirical distribution of the SNP--exposure associations,
it is also sensitive to SNP coding. 
In particular, 
data-driven all-positive recoding can further reduce 
${\hat{\theta}}_2$
and increase instability, 
as noted previously by Lin et al. (2022). 
In our implementation, 
we therefore adopt the fixed minor-allele-based coding scheme introduced above, 
using minor allele frequency (MAF) information when available, 
and examine alternative coding choices in the simulation study.

The measurement-error attenuation derived above motivates a simple correction:
subtracting the bias term from the denominator yields the debiased Egger (dEgger) estimator
$${\hat{\beta}}_{\rm dEgger}=\frac{{\hat{\theta}}_1}{{\hat{\theta}}_2-\Delta},$$
which removes the attenuation due to measurement error.
However, dEgger still relies on deterministic IV selection
and therefore remains subject to winner's curse bias
when the same exposure GWAS is used for both selection and estimation.
We thus regard dEgger as a conceptual intermediate step
and develop the rerandomized Egger estimator below.

\subsection{The REgger estimator}
Having identified the source of measurement-error bias in MR-Egger, 
we now extend the RIVW framework to jointly correct
measurement-error bias and winner's curse bias
while allowing more IVs to be included in the analysis.
To this end, 
we adopt the rerandomized IV-selection strategy of Ma et al. \cite{ma2023breaking}. 
For each SNP $j=1,2,\cdots,p$, 
generate independent noise $Z_j\sim N\left(0,\eta^2\right)$ 
and define the selected IV set as
$$S_\lambda=\left\{j:\left|\frac{{\hat{\gamma}}_j}{\sigma_{Xj}}+Z_j\right|>\lambda\right\}.$$
Let $p_\lambda=\left|S_\lambda\right|$ denote its size.
This randomization breaks the deterministic dependence 
between the selection event 
and the observed SNP--exposure effect estimate,
thereby mitigating winner's curse.
The tuning parameter $\eta$ controls the magnitude of the perturbation.
Define the average IV strength and effective sample size under random selection as  
$$\kappa_\lambda=p_\lambda^{-1}\sum_{j\in S_\lambda}{\sigma_{Xj}^{-2}}{\gamma_j^2}, \quad
\psi_\lambda=\kappa_\lambda\sqrt{p_\lambda}/\max(1,\lambda),$$
which govern the asymptotic behavior of the REgger estimator.
Conditional on this rerandomized selection, 
the Rao-Blackwellized (RB) estimator ${\hat{\gamma}}_{j,\rm RB}$
provides an unbiased estimator of $\gamma_j$
and attains the minimum variance among all unbiased estimators 
for a given $\eta$. 
Its variance is estimated by ${\hat{\sigma}}_{Xj,\rm RB}^2$
(see Appendix~\textcolor{blue}{S.2} for more details).

Since $\Delta$ captures the contribution of imprecise SNP--exposure estimates, 
we correct the denominator of the Egger estimator 
by subtracting its rerandomized counterpart $\Delta_\lambda$.
Specifically, we define
$${\hat{\theta}}_{1,\lambda}=\Big(\sum_{j\in S_\lambda}\sigma_{Yj}^{-2}\Big)\Big(\sum_{j\in S_\lambda}{\sigma_{Yj}^{-2}{{\hat{\gamma}}_{j,\rm RB}\hat{\Gamma}_j}}\Big)
  -\Big(\sum_{j\in S_\lambda}{\sigma_{Yj}^{-2}{\hat{\gamma}}_{j,\rm RB}}\Big)\Big(\sum_{j\in S_\lambda}{\sigma_{Yj}^{-2}{\hat{\Gamma}}_j}\Big),
$$
and
$${\hat{\theta}}_{2,\lambda}  = \Big(\sum_{j\in S_\lambda}\sigma_{Yj}^{-2}\Big)\Big(\sum_{j\in S_\lambda}{\sigma_{Yj}^{-2}{\hat{\gamma}}_{j,\rm RB}^2}\Big)
-\Big(\sum_{j\in S_\lambda}{\sigma_{Yj}^{-2}{\hat{\gamma}}_{j,\rm RB}}\Big)^2-\Delta_\lambda,
$$
where
$$\Delta_\lambda=\Big(\sum_{j\in S_\lambda}\sigma_{Yj}^{-2}\Big)
                 \Big(\sum_{j\in S_\lambda}{\sigma_{Yj}^{-2}{\hat{\sigma}}_{Xj,\rm RB}^2}\Big)
                -\sum_{j\in S_\lambda}{\sigma_{Yj}^{-4}{\hat{\sigma}}_{Xj,\rm RB}^2}.
$$
The rerandomized Egger (REgger) estimator is then given by
$$
  {\hat{\beta}}_{\rm RE}=\frac{{\hat{\theta}}_{1,\lambda}}{{\hat{\theta}}_{2,\lambda}}.$$
Together, 
rerandomized selection and measurement-error correction 
allow REgger to mitigate both winner's curse bias 
and measurement-error bias, 
while preserving the MR-Egger framework 
for accommodating directional pleiotropy.

\paragraph{Illustrative simulation.}
Here, we conduct a small benchmark simulation
to provide an intuitive illustration of the methodological gain of REgger.
The complete simulation framework is described in Section~4.1.
Detailed settings, including the coding-aligned data-generating mechanism, 
are provided in Appendix~\textcolor{blue}{S.1}.
We compare REgger with two MR-Egger analysis strategies:
unselected MR-Egger and MR-Egger after genome-wide significant selection at \(5\times10^{-8}\).
In this controlled setting, the former isolates measurement error bias, 
whereas the latter additionally reflects selection-induced winner's curse bias.

\begin{figure}[h] 
  \centering
  \includegraphics[width=1.0\textwidth]{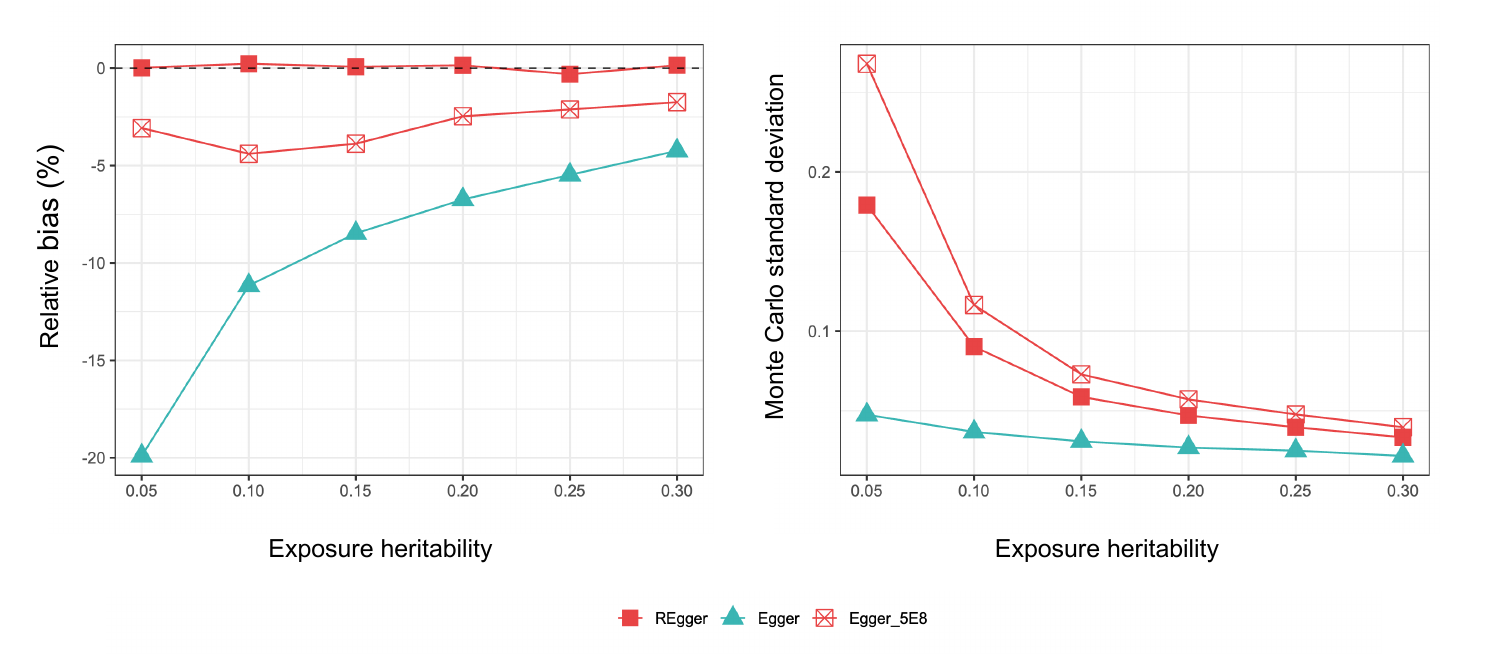}
\caption{
Illustrative simulation under directional pleiotropy with \(\beta=0.2\). 
Panel A shows relative bias, and Panel B shows the Monte Carlo standard deviation. 
Methods compared are REgger, unselected MR-Egger (Egger), and genome-wide-significant MR-Egger at \(5\times10^{-8}\) (Egger\_5E8). 
Exposure heritability is the total exposure variance explained by the simulated SNPs.
}
\label{module1_directional}
\end{figure}

Figure~\ref{module1_directional} shows the results under directional pleiotropy 
with a nonzero causal effect.
Unselected MR-Egger substantially underestimates the causal effect, 
especially when the exposure heritability explained by the SNPs is low.
Genome-wide significant selection reduces this attenuation, 
but the estimator remains underestimated because of winner's curse bias.
In contrast, REgger produces estimates centered close to the true causal effect 
across the examined levels of exposure heritability, 
indicating that rerandomized selection and Rao--Blackwellized correction 
effectively reduce both measurement error bias and winner's curse bias.

For variability, genome-wide significant MR-Egger has the largest Monte Carlo standard deviation, 
because strict selection discards a substantial amount of information.
REgger uses a more relaxed rerandomized selection procedure and retains more information, 
thereby showing lower variability than the strictly selected Egger estimator.
Although unselected MR-Egger has the smallest empirical variability, 
this reflects stable underestimation rather than valid estimation.
Similar patterns are observed under balanced pleiotropy 
(Appendix~\textcolor{blue}{S.1}).

\subsection{Construction of variance estimator}
A closed-form variance expression for ${\hat{\beta}}_{\rm RE}$ 
is not straightforward,
because
$\hat{\beta}_{\rm RE}$ is a ratio estimator 
whose numerator and denominator both depend on estimated quantities. 
To study its large-sample behavior, 
we rewrite
$${\hat{\beta}}_{\rm RE}-\beta
 =\frac{\theta_{2,\lambda}}{{\hat{\theta}}_{2,\lambda}}
 \cdot
 \frac{{\hat{\theta}}_{1,\lambda}-\beta{\hat{\theta}}_{2,\lambda}}{\theta_{2,\lambda}},$$
where 
$$ \theta_{2,\lambda}
	=\Big(\sum_{j\in S_\lambda}\sigma_{Yj}^{-2}\Big)\Big(\sum_{j\in S_\lambda}{\sigma_{Yj}^{-2}\gamma_j^2}\Big)
	-\Big(\sum_{j\in S_\lambda}{\sigma_{Yj}^{-2}\gamma_j}\Big)^2.
$$
Under regularity conditions, we show (see Appendix~\textcolor{blue}{S.3} for details) that
$${\hat{\theta}}_{2,\lambda}/{\theta_{2,\lambda}}\xrightarrow{P}1$$ 
as $\psi_\lambda \to \infty$ and $p_\lambda \to \infty$.
Therefore, the leading contribution to the variability of
${\hat{\beta}}_{\rm RE}$ 
comes from the term
 $\hat{\theta}_{1,\lambda}-\beta{\hat{\theta}}_{2,\lambda}$. 
We decompose it as follows (see Appendix~\textcolor{blue}{S.3} for details):
$${\hat{\theta}}_{1,\lambda}-\beta{\hat{\theta}}_{2,\lambda}
=A_\lambda+\Delta_{A_\lambda},$$ 
where 
$\Delta_{A_\lambda}$ 
is asymptotically negligible to $A_\lambda$, and
$$A_\lambda=\sum_{j\in S_\lambda}{\sigma_{Yj}^{-2}u_{j,\rm RE}},$$
with
$$ 
	u_{j,\rm RE}
	=\left({\hat{\gamma}}_{j,\rm RB}{\hat{\Gamma}}_j-\beta\left({\hat{\gamma}}_{j,\rm RB}^2-{\hat{\sigma}}_{Xj,\rm RB}^2\right)-\mu_\alpha{\hat{\gamma}}_{j,\rm RB}
   \right)\sum_{j\in S_\lambda}\sigma_{Yj}^{-2}
   -\left({\hat{\Gamma}}_j-\beta{\hat{\gamma}}_{j,\rm RB}-\mu_\alpha\right)\sum_{j\in S_\lambda}{\sigma_{Yj}^{-2}\gamma_j}.
$$
Conditional on the selected set $S_\lambda$, 
the variables $u_{j,{\rm RE}}$ are independent and have zero mean.
This motivates a sum-of-squares variance estimator for the leading term $A_\lambda$.
To obtain a feasible variance estimator, 
we replace the unknown parameters
$\beta$, $\mu_\alpha$, and $\gamma_j$ by their estimators.
In particular, 
analogous to MR-Egger regression, 
we define the corresponding intercept estimator on the selected set as
$${\hat{\mu}}_{\alpha,\lambda}
 =\frac{\sum_{j\in S_\lambda}{\sigma_{Yj}^{-2}\left({\hat{\Gamma}}_j-{\hat{\beta}}_{\rm RE}{\hat{\gamma}}_{j,\rm RB}\right)}}
       {\sum_{j\in S_\lambda}\sigma_{Yj}^{-2}}.
$$
Specifically, let
$$
  \hat{u}_{j,\rm RE} = 
 \Big({\hat{\gamma}}_{j,\rm RB}{\hat{\Gamma}}_j
		-{\hat{\beta}}_{\rm RE}\left({\hat{\gamma}}_{j,\rm RB}^2
		-{\hat{\sigma}}_{Xj,\rm RB}^2\right)-{\hat{\mu}}_{\alpha,\lambda}{\hat{\gamma}}_{j,\rm RB}\Big)\sum_{j\in S_\lambda}\sigma_{Yj}^{-2}
    -\Big({\hat{\Gamma}}_j-{\hat{\beta}}_{\rm RE}{\hat{\gamma}}_{j,\rm RB}-{\hat{\mu}}_{\alpha, \lambda}\Big)\sum_{j\in S_\lambda}{\sigma_{Yj}^{-2}\hat{\gamma}_{j, \rm RB}}
$$
and we define
$$ \hat{V}_{A_\lambda}
   =\sum_{j\in S_\lambda}{\sigma_{Yj}^{-4}\hat{u}_{j,\rm RE}^2}
$$
Accordingly, the feasible variance estimator for 
$\hat{\beta}_{\rm RE}$ is
$${\widehat{V}}_{\rm RE}
   =\frac{\hat{V}_{A_\lambda}}{{\hat{\theta}}_{2,\lambda}^2}.
$$
The explicit form of $\Delta_{A_\lambda}$
and additional technical details
are given in Appendix~\textcolor{blue}{S.3}. 
To establish the large-sample properties of REgger, 
we introduce the following additional assumptions.

\begin{assumption} \label{borrwo_RIVW} 
	The threshold satisfies: $\lambda\rightarrow\infty$. 
\end{assumption}

\begin{assumption} \label{NO_dominant_RB}
	The true SNP--exposure effects, $\{\gamma_j:j\in S_\lambda\}$, satisfy
	$$\frac{\max_{j\in S_\lambda}{\gamma_j^2}}{\sum_{j\in S_\lambda}\gamma_j^2}\rightarrow0.$$
\end{assumption}

Assumption \ref{borrwo_RIVW} is standard in the RIVW literature 
and imposes a diverging selection threshold 
to account for multiple testing in GWAS.
Assumption \ref{NO_dominant_RB} rules out a dominant IV in $S_\lambda$.

\begin{thm}
  Under Assumptions 
  \ref{Assume1}, 
  \ref{Assume2}, 
  \ref{InSIDE_Assumption}, 
  and \ref{borrwo_RIVW}, 
  as the effective sample size $\psi_\lambda\rightarrow\infty$ and $p_\lambda\rightarrow\infty$, 
  we have 
  $$\hat{\beta}_{\rm RE}\xrightarrow{P}\beta.$$
  If, in addition, Assumption \ref{NO_dominant_RB} holds, then
  $${\widehat{V}}_{\rm RE}^{-\frac{1}{2}}\left({\hat{\beta}}_{\rm RE}-\beta\right)\xrightarrow{D}N\left(0,1\right).$$
\end{thm}

\section{Simulations} \label{Simulation}
\subsection{Simulation settings}
Following common practice in the MR literature \cite{ye2021debiased,ma2023breaking}, 
we directly simulate the summary statistics 
 ${\hat{\gamma}}_j$ and ${\hat{\Gamma}}_j$  
 under Assumption \ref{Assume2}. 
Specifically,
$$\left[\begin{matrix}{\hat{\gamma}}_j\\{\hat{\Gamma}}_j\\\end{matrix}\right]\sim N\left(\left[\begin{matrix}\gamma_j\\\beta\gamma_j+\alpha_j\\\end{matrix}\right],\left[\begin{matrix}\sigma_{Xj}^2&0\\0&\sigma_{Yj}^2\\\end{matrix}\right]\right),
$$
where $\sigma_{Xj}={1}/{\sqrt{n_X}}$, $\sigma_{Yj}={1}/{\sqrt{n_Y}}$, and $n_X=n_Y=200{,}000$, 
reflecting typical GWAS sample sizes.
The true effects $(\gamma_j, \alpha_j)$ are generated from a four-component mixture distribution: 
$$\left[\begin{matrix}\gamma_j\\\alpha_j\\\end{matrix}\right]
   \sim\pi_1\left[\begin{matrix}N\left(\mu_\gamma,\varepsilon_x^2\right)\\\delta_0\\\end{matrix}\right]
   +\pi_2\left[\begin{matrix}N\left(\mu_\gamma,\varepsilon_x^2\right)\\N\left(\mu_\alpha,\varepsilon_\alpha^2\right)\\\end{matrix}\right]
   +\pi_3\left[\begin{matrix}\delta_0\\N\left(\mu_\alpha,\varepsilon_\alpha^2\right)\\\end{matrix}\right]
   +\left(1-\pi_1-\pi_2-\pi_3\right)\left[\begin{matrix}\delta_0\\\delta_0\\\end{matrix}\right],
$$
where $\delta_0$ denotes the Dirac measure centered at zero.
Here,
$\pi_1$ represents the proportion of valid IVs,
$\pi_2$ the proportion of pleiotropic IVs affecting both the exposure and the outcome, 
$\pi_3$ the proportion influencing only the outcome, 
and the remainder are null IVs.
The InSIDE assumption holds within the first three components by construction, 
whereas the presence of null IVs may induce mild finite-sample deviations when 
all SNPs are analyzed jointly.

Because all variables are standardized and SNPs are assumed independent, 
 the exposure and outcome heritabilities satisfy
\begin{align*}
  h_X^2=&{{\rm Var}\Big(\sum_{j=1}^{p}{\gamma_jG_j}\Big)}\Big/{{\rm Var}(X)}=\sum_{j=1}^{p}{\gamma_j^2{\rm Var}(G_j)}\approx p\left(\pi_1+\pi_2\right)\left(\mu_\gamma^2+\varepsilon_x^2\right),\\
  h_Y^2=&{{\rm Var}\Big(\sum_{j=1}^{p}{\Gamma_jG_j}\Big)}\Big/{{\rm Var}(Y)}=\sum_{j=1}^{p}\left(\beta\gamma_j+\alpha_j\right)^2{\rm Var}(G_j)\approx \beta^2h_X^2 +p\left(\pi_2+\pi_3\right)\left(\mu_\alpha^2+\varepsilon_\alpha^2\right).
\end{align*}
We set $\mu_\gamma=0.001$ and 
       $\varepsilon_x^2=\varepsilon_\alpha^2=0.0001$, with equal mixture proportions
       $\pi_1=\pi_2=\pi_3$. 
The value of $\pi_1$ varies from $0.001$ to $0.010$ in increments of $0.001$, 
 corresponding to exposure heritabilities from approximately $0.04$ to $0.40$. 
Two causal scenarios are examined: a non-null causal effect ($\beta = 0.2$) 
 and a null effect ($\beta = 0$),
 under two pleiotropy settings: directional pleiotropy ($\mu_\alpha = 0.005$) 
 and balanced pleiotropy ($\mu_\alpha = 0$).
For each configuration, we simulate $p=200{,}000$ independent SNPs 
 and conduct $1{,}000$ Monte Carlo repetitions.
Performance is evaluated using five metrics:
bias (reported as relative bias when $\beta\neq 0$ and as bias when $\beta=0$),
 Monte Carlo standard deviation (SD), 
 average estimated standard error (SE), 
 mean squared error (MSE), 
 and coverage probability (CP) of 95\% confidence intervals.

We compare our proposed estimators with
 dIVW \cite{ye2021debiased}, RIVW \cite{ma2023breaking}, 
 robust adjusted profile score (MR-RAPS) \cite{zhao2020statistical}, 
 and MR-Egger \cite{bowden2015mendelian}.
 These methods provide meaningful benchmarks for assessing performance under weak instruments and pleiotropy.
 For RIVW and REgger, instruments are selected using a relaxed significance threshold of $P<5\times{10}^{-5}$,
whereas all other methods use the conventional genome-wide threshold  $P<5\times{10}^{-8}$. 
In the random IV-selection step for both RIVW and REgger, the noise parameter is fixed at  $\eta=0.5$.
This unified framework ensures a fair comparison across methods while maintaining settings consistent with large-scale MR studies.

\subsection{Simulation results}
\begin{figure}[h] 
  \centering
  \includegraphics[width=1.0\textwidth]{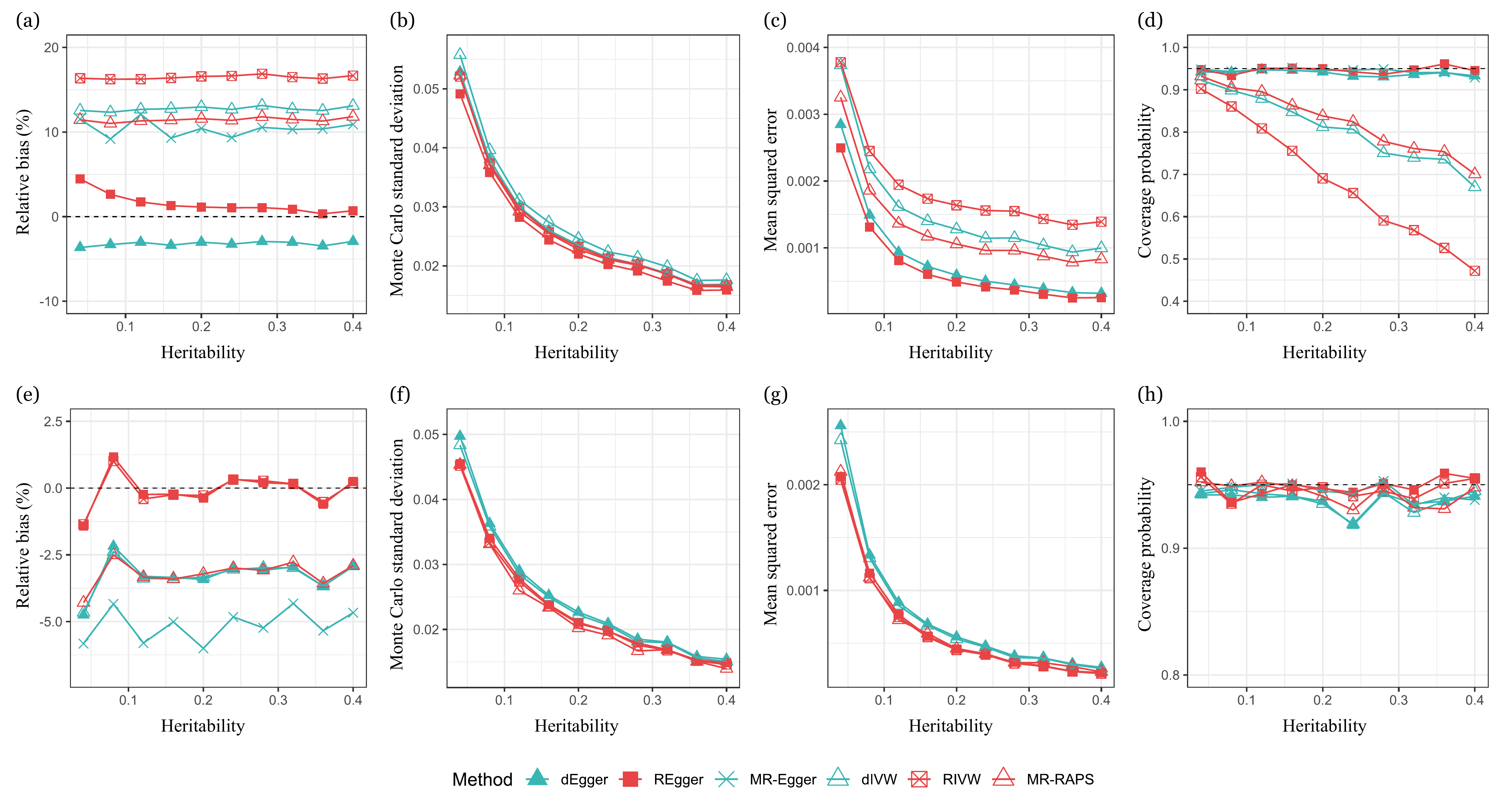}
  \caption{Relative bias, standard deviation, mean squared error, and coverage probability
   of the REgger estimator and several competing MR methods versus the heritability of SNPs on the exposure.
   The top row corresponds to directional pleiotropy with simulation settings
    $\mu_\gamma=0.001$,  $\beta=0.2$, $\mu_\alpha=0.005$,
   and $\pi_1=\pi_2=\pi_3$ varying from $0.001$ to $0.010$ in increments of $0.001$.
  The bottom row corresponds to balanced pleiotropy with the same settings except that $\mu_\alpha = 0$.}
  \label{sim_main_beta02}
\end{figure}

Figure \ref{sim_main_beta02} summarizes the performance of all methods under a non-null causal effect ($\beta=0.2$), 
with relative bias, SD, MSE, and CP plotted against exposure heritability. 
Since exposure heritability increases approximately linearly with $\pi_1+\pi_2$,
the proportion of variants associated with the exposure,
this x-axis effectively indexes overall instrument strength.

The top row corresponds to directional pleiotropy ($\mu_\alpha = 0.005$),
 whereas the bottom row represents balanced pleiotropy ($\mu_\alpha = 0$).
Under directional pleiotropy, 
 all competing estimators display notable relative bias (panel a), often exceeding 10\%.
This bias arises from two opposing sources: winner's curse, which biases estimates downward, 
and directional pleiotropy, which shifts them toward the direction of pleiotropy (upward in our simulations).
As a result,  
 the observed bias pattern reflects a combination of underestimation and overestimation across methods.
In contrast, the proposed REgger estimator exhibits minimal bias that converges toward zero 
as exposure heritability increases, owing to its joint correction for both winner's curse 
and directional pleiotropy. 
For reference, 
we also report the dEgger estimator (Section~3.1),
which corrects measurement-error bias in Egger regression but does not address selection-induced 
winner's curse; accordingly, it tends to underestimate the causal effect under these settings.
As shown in panel b, 
 REgger also achieves the smallest empirical SD by efficiently incorporating weak instruments, 
  thereby improving precision.
Across all methods, 
the average estimated SE closely tracks the empirical SD on the log--log scale (Figure \textcolor{blue}{S2}), 
and the same agreement is observed in the additional simulation scenarios.
In terms of MSE (panel c),
REgger attains the smallest values, combining low bias with reduced variability.
It also achieves coverage probabilities close to the nominal 95\%, whereas competing methods tend to under-cover because of biased point estimates.
Under balanced pleiotropy (second row), 
where directional bias is removed, 
all methods exhibit mild bias toward the null (panel e), 
primarily due to the winner's curse. 
In this setting, REgger performs comparably to RIVW, 
with both showing minimal bias, relatively small SDs, and near-nominal coverage probabilities.

\begin{figure}[h]
  \centering
  \includegraphics[width=0.95\textwidth]{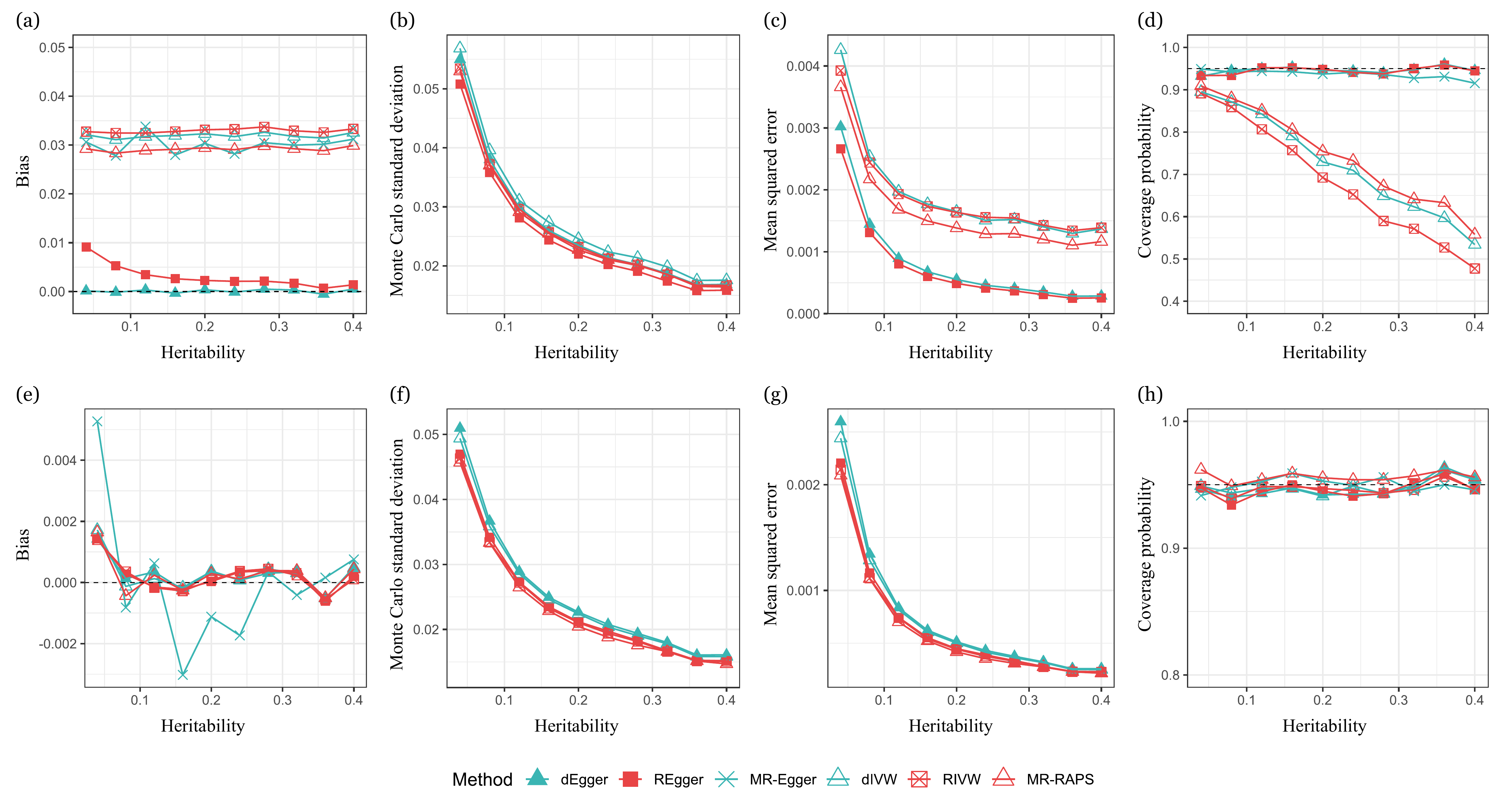} 
  \caption{Bias, standard deviation, mean squared error, and coverage probability
   		   of the REgger estimator and several competing MR methods versus 
		   the heritability of SNPs on the exposure.
   		   The top row corresponds to directional pleiotropy with simulation settings
		   $\mu_\gamma=0.001$,  $\beta=0$, $\mu_\alpha=0.005$, 
		   and $\pi_1=\pi_2=\pi_3$ varying from $0.001$ to $0.010$ in increments of $0.001$.
		   The bottom row corresponds to balanced pleiotropy with the same settings except that $\mu_\alpha = 0$.}
  \label{sim_main_beta00}
\end{figure}

When the causal effect is null ($\beta=0$; Figure \ref{sim_main_beta00}), 
 the winner's curse is eliminated, 
 leaving directional pleiotropy as the only potential source of bias. 
Accordingly, the dEgger estimator, which was biased under selection when $\beta\neq0$,
now becomes essentially unbiased under directional pleiotropy (first row of Figure \ref{sim_main_beta00}).
All methods are nearly unbiased under balanced pleiotropy (second row), where neither source of bias is present.
The corresponding coverage probabilities in the second row are close to the nominal 95\%, 
and the remaining patterns are consistent with those shown in Figure \ref{sim_main_beta02}.

Finally, a supplementary scenario with $\mu_\gamma=0$ (Figures \textcolor{blue}{S3--S4})
further disentangles the sources of bias.
This setting satisfies $\sum_{j=1}^{p}{\sigma_{Yj}^{-2}\alpha_j\gamma_j}=0$,
under which RIVW remains unbiased. 
In this case, REgger attains a slightly smaller SD than RIVW, 
 while other methods exhibit bias toward the null, 
 indicating that the winner's curse is the dominant bias component in this configuration.
However, the assumption $\mu_\gamma=0$ is rather restrictive and unlikely to hold exactly 
 in practical applications.

\subsection{Tuning parameter selection}
We further vary the tuning parameter $\eta$ over the grid $(0.2,0.4,0.5,0.6,0.8)$
to assess the sensitivity of REgger to this choice.
The simulation settings are otherwise identical to those in Figure \ref{sim_main_beta02}.
We consider both directional and balanced pleiotropy scenarios by setting $\mu_\alpha=0.005$ 
and $\mu_\alpha=0$, respectively. 
Figure \textcolor{blue}{S5} summarizes the results, 
with the first row corresponding to directional pleiotropy and the second row to balanced pleiotropy.
Overall, REgger exhibits stable performance across a reasonable range of $\eta$, 
with particularly consistent behavior under balanced pleiotropy.
Based on these findings, 
we recommend using $\eta=0.5$ as a default in applications, 
matching the value used in RIVW under the same rerandomized IV-selection mechanism \cite{ma2023breaking}. 
This choice avoids dataset-specific tuning while delivering reliable performance in our simulations.

\subsection{Additional analyses on SNP coding and inference stability}
\begin{figure}[t] 
  \centering
  \includegraphics[width=1.0\textwidth]{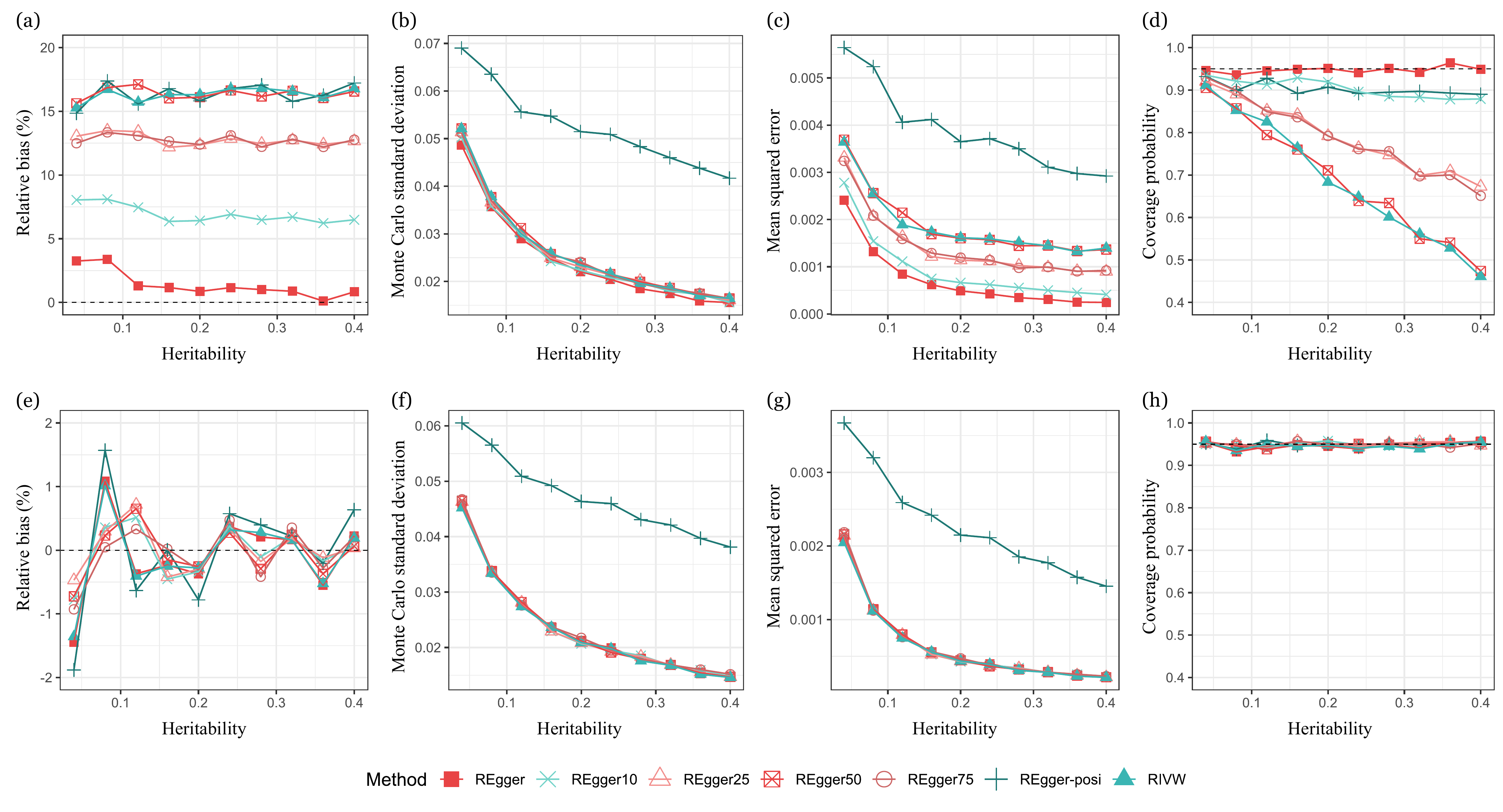}
  \caption{Relative bias, Monte Carlo standard deviation, mean squared error, 
   and coverage probability of the REgger estimator for different proportions of SNP sign flipping 
   (REgger10, REgger25, REgger50, REgger75 correspond to 10\%, 25\%, 50\% and 75\% of SNPs flipped, respectively, 
   REgger-posi denotes the conventional all-positive recoding scheme). 
   The first row corresponds to directional pleiotropy with simulation settings
    $\mu_\gamma=0.001$,  $\beta=0.2$, $\mu_\alpha=0.005$,
   and $\pi_1=\pi_2=\pi_3$ varying from $0.001$ to $0.010$ in increments of $0.001$;
  The second row corresponds to balanced pleiotropy with the same settings except that $\mu_\alpha = 0$.}
  \label{sim_robust_beta02}
\end{figure}

We assess the sensitivity of REgger to potential violations of the InSIDE assumption by randomly flipping SNP signs at prespecified proportions and by comparing with the conventional all-positive recoding scheme (Figure \ref{sim_robust_beta02}) \cite{lin2022practical}.
Under balanced pleiotropy (bottom row), performance is broadly comparable across flip settings, whereas all-positive recoding yields noticeably larger SDs and less stable point estimates.
Under directional pleiotropy (top row), REgger's bias increases with the flip proportion and becomes comparable to RIVW around 50\% flipping; all-positive recoding again produces the largest SD, with bias similar to RIVW.
The null-effect results display the same qualitative pattern (Figure \textcolor{blue}{S6}).
Additional sensitivity scenarios are reported in Supplementary \textcolor{blue}{S.7.2}.

In an additional Rao--Blackwell-based effective-sample-size 
(RB-based ESS) simulation \textcolor{blue}{(Supplementary Section~S.7.3; Figure~S9--S10)}, 
REgger's finite-sample behavior stabilizes once the RB-based ESS exceeds approximately $20$
and the selected IV count exceeds approximately $200$; 
we use these thresholds as pragmatic guidelines in applications.

\section{Real data analysis} \label{Real}
We illustrate the practical performance of REgger using two real-data applications.
In a same-trait setting, 
where the RIVW estimator is theoretically optimal, 
we assess whether REgger attains comparable estimates without appreciable loss of efficiency. 
We then investigate the potential causal effects of 
size-specific high-density lipoprotein cholesterol (HDL-C) subfractions 
on non-cerebrocardiovascular atherosclerosis, 
to demonstrate the ability of REgger to recover biologically meaningful associations 
in a more complex etiologic setting.

\subsection{Data formation and harmonization}
We apply a two-stage preprocessing pipeline to the GWAS summary statistics: 
quality control (QC) and data harmonization.
During QC, we retain only SNPs included in the  HapMap 3 reference panel, 
exclude variants with minor allele frequency (MAF)  $< 0.01$, 
and remove SNPs located in the complex major histocompatibility region on Chromosome 6 (26Mb--34Mb).
In the harmonization stage, 
we first remove exposure SNPs that are unavailable in the outcome GWAS summary statistics.
We then identify independent SNPs by ensuring that no pair has 
linkage disequilibrium (LD) with $r^2>0.001$ 
within a $10$ Mb window. 
To avoid the selection bias associated with conventional $P$-value-based LD clumping \cite{hu2022mendelian},
we adopt the sigma-based LD clumping algorithm of Ma et al. \cite{ma2023breaking},
which prioritizes variants according to their standard errors
 rather than test statistics, 
thereby preserving the integrity of effect-size estimates while maintaining LD independence.
Finally, using allele frequency information, 
we resolve strand-ambiguous SNPs and harmonize effect alleles across the exposure and outcome datasets.

We apply the same MR methods and IV selection strategies as in the simulation study. 
For the real-data analyses, 
we focus on reporting the final
practically recommended procedures and widely used benchmarks; 
therefore, we do not include dIVW or dEgger, 
which are primarily used as intermediate reference estimators in our methodological development.
Specifically, RIVW and REgger use a relaxed significance threshold of 
$5\times{10}^{-5}$ to balance statistical power and weak instrument bias, 
whereas all other methods use the conventional genome-wide threshold 
($\lambda=5.45$, equivalent to $P<5\times{10}^{-8}$). 
For the methods employing rerandomized IV selection, 
we set $\eta=0.5$, consistent with the simulation setup. 

\subsection{Same trait analysis}
\begin{figure}[htbp]
  \centering
  \includegraphics[width=1.0\textwidth]{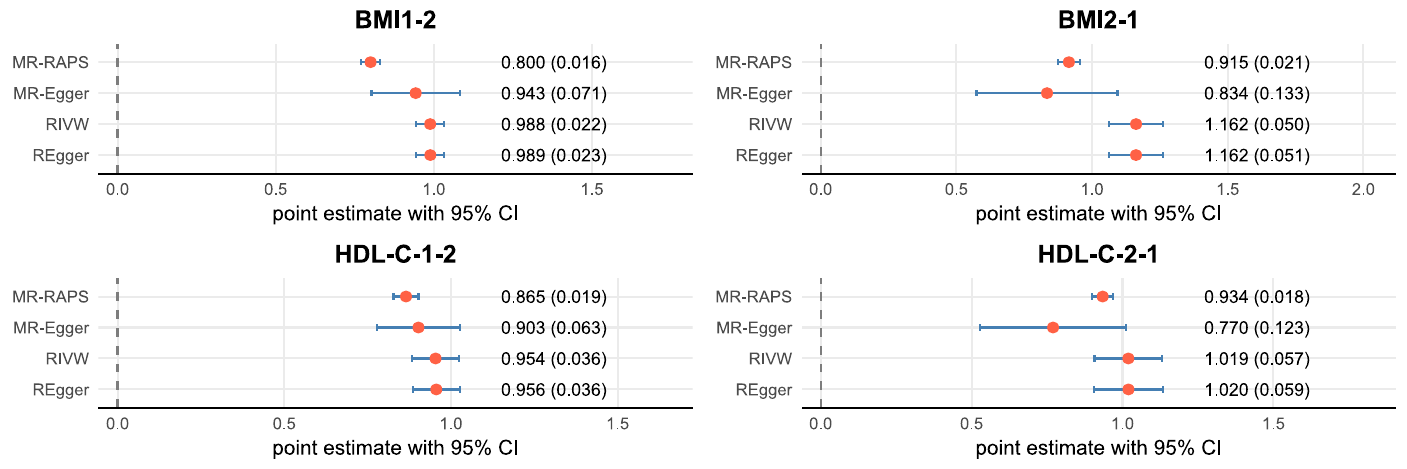}
  \caption{Same-trait results for different methods under sigma-based LD clumping}
  \label{fig_sametrait}
\end{figure}

In the same-trait analysis, we use two independent GWAS datasets for each trait
 as reciprocal exposure--outcome pairs, 
so that the true causal effect equals 1 by construction in the absence of pleiotropy \cite{zhao2020statistical,ma2023breaking}.
This design allows us to empirically assess the finite-sample performance 
of different estimators.
Using the same datasets as in previous RIVW work (see Table \textcolor{blue}{S2} for details), 
we analyze body mass index (BMI1 vs.\ BMI2) and high-density lipoprotein cholesterol (HDL-C1 vs.\ HDL-C2) \cite{ma2023breaking}. 
The four pairwise analyses (BMI1--2, BMI2--1, HDL-C1--2, and HDL-C2--1)
are shown in Figure \ref{fig_sametrait} 
with point estimates and 95\% confidence intervals. 

Across all comparisons, 
RIVW provides estimates closest to 1 with relatively narrow confidence intervals,
consistent with its theoretical efficiency when model assumptions hold.
REgger yields very similar point estimates with only slightly wider intervals,
reflecting the additional uncertainty from estimating the pleiotropy intercept 
while preserving good calibration.
MR-Egger yields more variable estimates and the widest confidence intervals, consistent with its known sensitivity to weak instruments and to all-positive recoding of SNP--exposure associations \cite{bowden2016assessing,lin2022practical}.
MR-RAPS systematically underestimates the causal effect (with estimates below $1$),
suggesting residual selection bias (winner's curse) despite the stringent IV threshold ($P<5\times{10}^{-8}$),
a pattern consistent with our simulation findings.
Under standard LD clumping (Figure \textcolor{blue}{S11}), 
both RIVW and REgger estimates are shifted below $1$, 
illustrating the impact of clumping-induced selection bias.  
More detailed summary results for REgger in the same-trait analyses are reported in 
 Table \textcolor{blue}{S3}.

\subsection{Size-specific HDL-C and non-cerebrocardiovascular atherosclerosis}
\begin{figure}[htbp]
  \centering
  \includegraphics[width=1.0\textwidth]{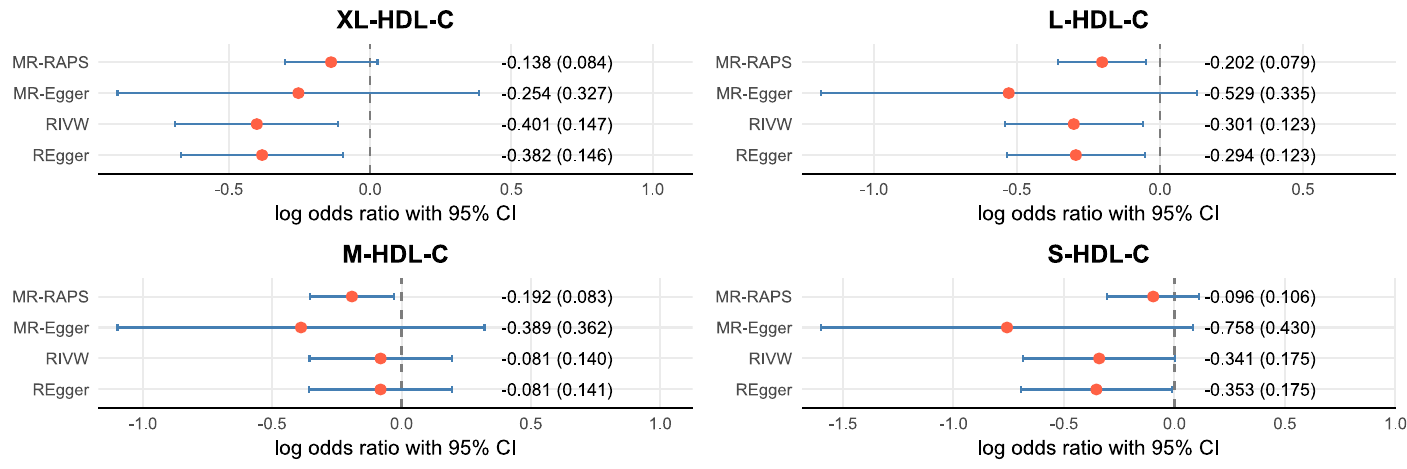}
  \caption{Causal effect estimates of size-specific HDL-C on non-cerebrocardiovascular atherosclerosis for different methods under sigma-based LD clumping.}
  \label{fig_hdl}
\end{figure}

Atherosclerosis, 
the most prevalent form of cardiovascular disease, 
arises from lipid accumulation and inflammation in large arteries and can 
lead to myocardial infarction, stroke, and impaired perfusion in other vascular beds, 
including peripheral artery disease 
and renovascular hypertension \cite{fan2022ather0,dai2025combination}. 
Although HDL-C is traditionally viewed as protective against atherosclerosis, 
clinical trials that simply increase total HDL-C have yielded inconsistent results \cite{casula2021hdl}. 
One plausible explanation is that HDL particles are heterogeneous in size, lipid content, 
and protein composition,
and that specific subfractions rather than total HDL-C may drive vascular protection \cite{casula2021hdl, endo2023hdl}.
To disentangle these effects, 
we conduct a two-sample MR analysis of four HDL-C subfractions,
very large (XL-HDL-C), large (L-HDL-C), medium (M-HDL-C), and small (S-HDL-C),
with non-cerebrocardiovascular atherosclerosis as the outcome,
which captures systemic atherosclerotic disease beyond the coronary and cerebrovascular beds.
Exposure GWASs for each subfraction are obtained from $115{,}082$ UK Biobank participants, 
in whom targeted high-throughput NMR metabolomics quantified cholesterol in each size class \cite{richardson2022char}. 
The outcome GWAS for non-cerebral, non-coronary atherosclerosis is taken from
 FinnGen release R12 ($19{,}807$ cases and $463{,}106$ controls) \cite{Kurki2023}. 
We apply the same set of methods and parameter choices as in the same-trait analysis.
The causal estimates are displayed in Figure \ref{fig_hdl}.

All four HDL-C subfractions yield negative point estimates, 
broadly consistent with observational evidence 
that higher HDL functionality is associated with lower atherosclerotic risk \cite{casula2021hdl,endo2023hdl}.
XL-HDL-C displays the strongest protective effect, 
with both REgger and RIVW yielding the largest significant negative point estimates. 
Moreover, the point estimates from the two methods differ slightly, 
 suggesting sensitivity to directional pleiotropy or related model deviations.
In addition, REgger shows a slightly smaller standard error than RIVW, 
 consistent with the simulation-based advantage of REgger when directional pleiotropy is present.  
L-HDL-C also demonstrates a protective association, 
for which RIVW and REgger again agree closely and achieve significance, 
whereas methods relying on more stringent fixed instrument thresholds 
tend to attenuate the effect toward the null.
For M-HDL-C and S-HDL-C, 
the evidence is weaker and less consistent across methods and clumping strategies.
Sensitivity checks using standard LD clumping (Figure \textcolor{blue}{S12}) 
tend to shift estimates toward the null, 
echoing the clumping-induced attenuation observed in the same-trait analyses 
and reinforcing the importance of careful IV selection.

Taken together, 
these results support XL-HDL-C and L-HDL-C as the HDL-C subfractions 
most plausibly contributing to protection against systemic (non-cerebrocardiovascular) atherosclerosis,
whereas medium and small HDL-C appear less clearly implicated.
From a clinical perspective,
 this pattern suggests that interventions preferentially enhancing very large and large HDL particles,
 rather than nonspecific elevation of total HDL-C,
 may represent a more promising strategy for reducing the burden of 
 generalized atherosclerotic disease.

\section{Discussion}
Mendelian randomization analyses frequently employ MR-Egger method as a robust alternative 
to the IVW method when directional pleiotropy is suspected \cite{bowden2015mendelian}. 
However, the original Egger regression ignores measurement error in the SNP--exposure associations, 
leading to attenuation of the causal estimate in the presence of many weak instruments,
and its inflated standard errors further erode statistical power \cite{bowden2016assessing,bowden2019improving}.
In this study,
we explicitly decomposed the weak-instrument bias of MR-Egger,
showed that it arises from inflation of the regression denominator due to measurement error,
and proposed a simple measurement-error correction (dEgger) as a conceptual intermediate step.
Building on this bias-correction insight, we then constructed the rerandomized Egger (REgger) estimator by combining Egger regression with a rerandomized IV-selection procedure and Rao--Blackwellized effect estimates.
Given the difficulty of analytically deriving variances for these ratio-type estimators, 
we developed regression-residual-based variance estimators 
that discard higher-order terms while retaining
the leading stochastic components.

Across a broad range of scenarios, 
our simulations indicate that the REgger estimator is well calibrated in realistic settings:
when the RB-based effective sample size exceeds approximately $20$ 
and at least about $200$ instruments are retained after selection, 
the estimated standard errors closely match the empirical Monte Carlo variability, 
and the REgger estimator exhibits near-normal sampling behavior.
Relative to the measurement-error-only correction, 
REgger additionally mitigates winner's curse bias and stabilizes inference after IV selection.
Compared with RIVW, REgger offers two main advantages.
First, under directional pleiotropy, 
 it provides robust bias correction and may achieve a slightly smaller standard error than RIVW.
Second, under balanced pleiotropy, 
 its performance is nearly identical to that of RIVW,
 differing mainly by slightly wider confidence intervals due to the extra intercept parameter.
These findings are further supported by our real-data analyses.
Taken together, 
 our results support REgger as a practical complement to RIVW in routine MR sensitivity analyses, 
 particularly when directional pleiotropy and weak instruments are of concern.

Several limitations warrant consideration. 
First, because SNP orientation simultaneously affects 
 the InSIDE condition and the stability of MR-Egger-type estimators,
 we adopt minor-allele coding rather than the conventional practice of 
 forcing all SNP--exposure estimates to be positive.
The all-positive convention can compress variability in $\hat{\gamma}_j$, 
 inflate the variance of ratio-type estimators,
 and, as our simulations suggest, 
 induce spurious correlation between $\gamma_j$ and $\alpha_j$, 
 thereby compromising InSIDE.
Minor-allele coding avoids effect-size-driven sign flipping
and does not materially reduce efficiency under balanced pleiotropy,
but it does not eliminate the fundamental fact 
that InSIDE is coding-dependent and remains untestable \cite{lin2022practical,burgess2017interpreting}.
Our simulations indicate that REgger is reasonably robust under moderate deviations from InSIDE,
especially compared with conventional MR-Egger,
yet its consistency still relies on InSIDE holding at least approximately for the selected instruments,
and further theoretical work on severe or structured violations would be valuable.
Second, our framework currently assumes independent instruments and no sample overlap. 
Future extensions should consider correlated pleiotropy, overlapping samples, 
and multivariable Egger analysis to broaden the method's applicability.

In conclusion, 
the REgger estimator extends the MR-Egger toolkit by 
jointly addressing weak-instrument bias, winner's curse bias, 
and coding-related instability, 
while retaining robustness to directional pleiotropy 
and providing a coherent variance estimation strategy.
It is straightforward to implement and performs well under realistic simulation and real-data settings, 
making it a useful and implementable addition to contemporary Mendelian randomization practice.

\section*{Acknowledgement}

\bibliographystyle{unsrtnat}  
\bibliography{reference}

\clearpage

\end{document}